\documentclass[fleqn,a4paper,12pt]{article}
\usepackage[a4paper,bindingoffset=0.2in,%
            left=0.3in,right=0.35in,top=1.5in,bottom=1.2in,%
            footskip=.35in]{geometry}
\usepackage{amsmath}
\usepackage{amssymb}
\usepackage{comment}
\usepackage{hyphenat}
\usepackage{multicol}
\usepackage{footnote}
\usepackage[hang,flushmargin]{footmisc}	

\newcommand{\qed}{\hfill$\Box$}

\usepackage[]{hyperref}
\hypersetup{
  colorlinks   = true, 
  urlcolor     = blue, 
  linkcolor    = blue, 
  citecolor   = blue 
}
\usepackage{subfigure}
\usepackage{graphicx}
\usepackage[margin=1cm]{caption}

\title{\textbf{Numerical studies on effect of operating conditions and geometrical components on the timing response and efficiency of RPC}}
\date{}
\begin{document}
\pagenumbering{gobble}
\maketitle
\vspace{-1.5cm}
\begin{center}
\large
A. Jash$^{a,c}$\footnotemark, N. Majumdar$^{a,c}$, S. Mukhopadhyay$^{a,c}$, S. Chattopadhyay$^{b,c}$
\\[0.6cm]
\small
\emph{
$^{a}$Applied Nuclear Physics Division, Saha Institute of Nuclear Physics, Block AF, Sector 1, Bidhannagar, Kolkata 700064, India
\\
$^{b}$Experimental High Energy Physics Division, Variable Energy Cyclotron Centre, Block AF, Sector 1, Bidhannagar, Kolkata 700064, India
\\
$^{c}$HBNI, Training School Complex, Anushakti Nagar, Mumbai 400094, India
}
\\[0.4cm]
E-mail: \href{mailto:abhik.jash@niser.ac.in}{abhik.jash@niser.ac.in}
\end{center}
\footnotetext{
\textbf{Present address}: Department of Particle Physics and Astrophysics, Weizmann Institute of
Science, Rehovot 7610001, Israel (E-mail: abhik.jash@weizmann.ac.il).
}

%
\vspace{1cm}
\abstract{
\noindent
A detailed numerical simulation has been performed to investigate the timing properties of a 
standard RPC geometry in order to optimize its application in INO-ICAL experiment designed for 
studying atmospheric neutrinos. The timing information provided by the RPC is an important 
observable in the experiment to determine the direction of the neutrinos. Effects of various 
operating conditions and design components on the RPC timing response, mainly the average
signal arrival time and the time resolution have been studied in this context.
The dependence of the detector efficiency on the same parameters has also been presented.
}
%
%
\newpage
\pagenumbering{arabic}
\begin{center}
\rule{\linewidth}{0.1pt}
\end{center}
\tableofcontents
\begin{center}
\rule{\linewidth}{0.1pt}
\end{center}
\section{Introduction}
\label{section:introduction}
Resistive Plate Chamber (RPC)~\cite{RPC} has been chosen as the active detector element for 
the magnetized Iron CALorimeter (ICAL) setup~\cite{ICAL} of 50 kTon target mass in the proposed 
underground laboratory facility of India-based Neutrino Observatory (INO). The ICAL has been 
designed for detection of the muons propagating through the calorimeter, identification of their charges
and accurate determination of their energies and directions. These will allow for a sensitivity to 
the neutrino mass hierarchy and precision measurement of the neutrino mixing parameters. To 
accomplish the task, the ICAL will be equipped with RPCs stacked in horizontal layers interleaving 
151 layers of iron plates of thickness 5.6 cm which will be magnetized to produce a uniform 
magnetic field of 1.3 T over the entire plate thickness. Muons of opposite charges ($\mu^{-}$, 
$\mu^{+}$) produced by the charged current interaction of incoming atmospheric muon 
neutrinos ($\nu_{\mu}$, {$\bar{\nu}_{\mu}$) with the iron nuclei in the target mass will bend 
differently under the action of the magnetic field and their path will be tracked using the two 
dimensional position information provided by each RPC layer. In addition, the timing response
of the RPCs will allow for a determination of directionality of the muons by distinguishing between
up-going and down-going muons. All these measurements will allow to probe several important observables, 
such as, the earth-matter effect of $\nu_{\mu}$ and $\bar{\nu}_{\mu}$ that travel through the
earth before reaching the ICAL, their path length and the momenta. Fast timing response with 
time resolution of the order of one nanosecond  and position resolution of the order of one 
centimeter from each detector layer are the basic requirements to achieve these goals.
\\
A thorough understanding of the operation of RPC and dependence of its performance on different
conditions will help us to optimize its application in the ICAL as well as interpret and predict the
experimental results. The RPC made up of float glass electrodes has been opted for muon 
detection in ICAL. However, a simultaneous R\&D has been underway to investigate the feasibility
of using bakelite electrodes which can offer several advantages like high rate capability, easy 
handling and fabrication and can be an alternative option for the ICAL.  
In the present work, a detailed numerical simulation has been performed for this purpose to 
investigate the dynamics of a bakelite RPC for different operational as well as geometrical 
parameters which may be useful for validating its choice. Also, the numerical method has 
been used to predict the behavior of the detector at conditions which are difficult to 
realize using experiments.
In this context, computation of the entire working procedure of an RPC beginning from the 
primary ionization caused by the passage of muons through it, till the generation of signal on its
read-outs, has been performed using a simulation framework developed for the gaseous detectors.
The main thrust has been put on the investigation of its timing performance under various 
conditions. A straightforward approach based on the basic principles of electronic signal detection 
has been implemented to calculate the timing response in the process of simulating 
the induced current signal.
The same simulation has been performed for computing the timing properties of a glass RPC
for which experimental data are available~\cite{manas_glassRPCDevelopment} 
to compare with the numerical work. In this paper, the material property (bakelite or glass)
has been used for calculating the electric field and weighting field within the RPC gas chamber which
depends on the dielectric constant of the chosen material.
\\
The content of the paper has been arranged in the following manner. The scheme of numerical simulation
adopted to calculate the RPC response has been described in section~\ref{section:simulationScheme}.
In section~\ref{section:comp_simu_expt} the simulated results for a glass RPC has been compared 
with the available experimental data. 
The effect of different operating conditions and the geometrical components on the timing properties 
and the efficiency of a bakelite RPC, as found from simulation, has been presented in 
section~\ref{section:numericalResult}. The computer resources used for the present simulation 
work have been mentioned in section~\ref{section:resources}. Finally, concluding remarks are 
made in section~\ref{section:Conclusion}.
\section{Simulation scheme}
\label{section:simulationScheme}
RPC being a gaseous detector, works on the principle of ionization phenomenon caused by the passage of a 
charged particle/radiation through its gaseous volume. When an energetic charged particle or 
radiation passes through the gaseous medium, it imparts its energy partially to the molecules of
the gas mixture, causing their ionization or excitation. In presence of the electric field, the electrons 
and cations, created in the primary ionization, move towards the anode and the cathode, 
respectively. At lower values of the electric field, the electrons and cations may get lost due to 
processes like recombination, attachment, electron capture etc. When the electric field is sufficiently 
high, the primary electrons may gain enough kinetic energy to ionize other molecules when they collide with 
them, liberating more electrons and cations. Depending upon the kinetic energy gathered by the 
primaries, they can cause further ionization even after the secondary level. On the other hand, 
the cations, being 
very heavy, generally move slower and the probability of secondary ionization by the cations is 
quite less usually. The movement of all the ions, thus produced in a cascade of ionization, commonly 
known as avalanche, induces a current on the conductive read-out panel. 
Following the Shockley-Ramo theorem~\cite{paper_Shockley, paper_Ramo}, the current induced
by a charge $q$, moving with an instantaneous velocity $\overrightarrow{v}(t)$, is given by 
the equation \eqref{equ:shockley_ramo}:
\begin{equation}
i(t) = q\overrightarrow{v}(t).\overrightarrow{W}(\overrightarrow{x}(t)) 
\label{equ:shockley_ramo}
\end{equation}
where, $\overrightarrow{W}(\overrightarrow{x}(t))$ is the weighting field for a read-out and that
is calculated as the electric field produced at the instantaneous position ($\overrightarrow{x}(t)$)
of the charge ($q$) when the read-out of interest is kept
at unit potential while all other available conductors are grounded. The 
signal thus collected from the read-out strips is the characteristic response of the RPC that depends 
on its geometry, electric field and the filling gas mixture for a given incident particle or radiation. 
In the present work, the timing properties of a bakelite RPC have been studied for variation in 
its operational parameters, such as, supplied voltage and filling gas mixture. In addition, the effect
of geometrical components, such as, the edge and the button spacer has been 
investigated as well. To accomplish these, numerical simulation of the RPC signal for different 
operational and geometrical conditions has been performed by calculating the current induced 
on one of its read-out strips using Garfield~\cite{Garfield}. Garfield is a simulation framework which 
has interfaces to several toolkits, namely, HEED~\cite{HEED}, neBEM~\cite{neBEM}, Magboltz~\cite{Magboltz},
to compute different auxiliary components relevant to the gaseous detector dynamics, such as, primary 
ionization, electrostatic field, transport properties of the gas respectively. The final stage of signal 
calculation is computed by the Garfield with the use of relevant data produced by the respective 
toolkits.
\\
The information on the primary ionization in the RPC gas chamber filled with a given gas mixture
due to the passage of a muon with a specific energy and direction can be obtained using HEED.
\begin{figure}[!htb]
 \centering
  \subfigure[]{
    \includegraphics[width=0.42\textwidth, trim={0.1cm 0.1cm 2.2cm 2.2cm},clip]{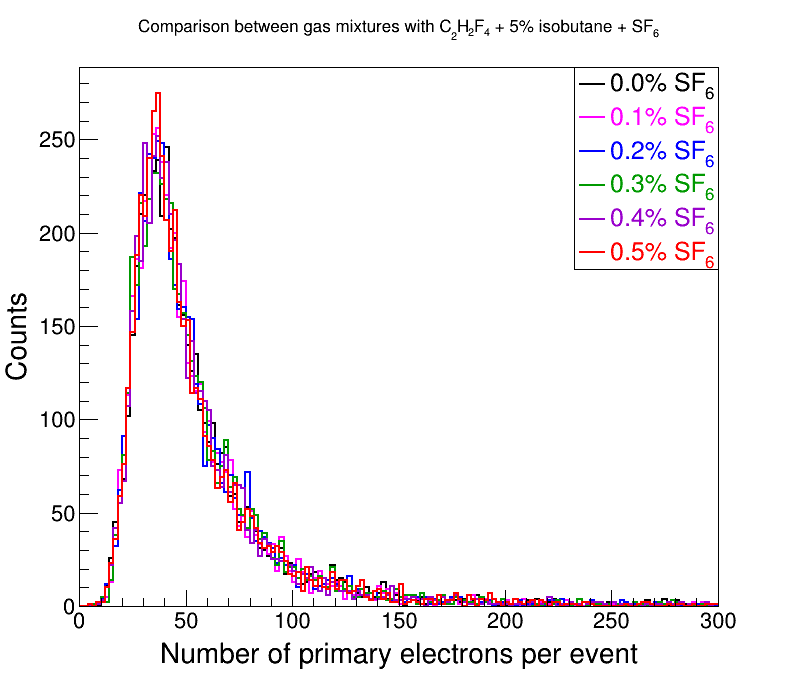}
  \label{fig:nClusters_HEED}
  }
  \subfigure[]{
     \includegraphics[width=0.42\textwidth, trim={0.1cm 0.1cm 2.2cm 2.2cm},clip]{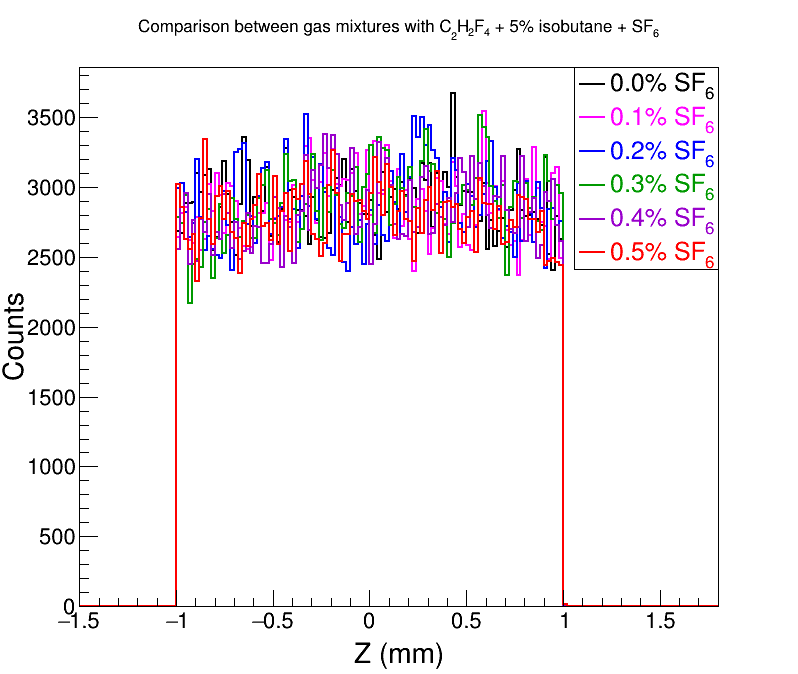}
\label{fig:clusterDist_HEED}
  }
\caption{\subref{fig:nClusters_HEED} Distribution of number of primary electrons per event and 
\subref{fig:clusterDist_HEED} spatial distribution of primaries in a 2 mm gas gap when 5000 
mono-energetic muons, each of energy 2 GeV traveling in the same direction pass through the 
chamber containing the mixture C$_{2}$H$_{2}$F$_{4}$, 5\% i-C$_{4}$H$_{10}$ and different 
fractions of SF$_{6}$.}
\label{fig:primaryIonization_HEED}
\end{figure}
As an example, figure~\ref{fig:nClusters_HEED} shows a distribution of number of primary electrons 
generated per event by the passage of 5000 muons each of energy 2 GeV traveling in a fixed direction 
(5$^{\circ}$ with respect to the zenith) through a 2 mm gas gap containing C$_{2}$H$_{2}$F$_{4}$,
5\% i-C$_{4}$H$_{10}$ and different fractions of SF$_{6}$. 
Landau fit of the distribution produces a most probable value (MPV) of 34.81 denoting a primary electron 
density of 174.05$/$cm for the given gas mixture. Figure~\ref{fig:clusterDist_HEED} shows the 
spatial distribution of the primary electrons along the thickness of the gas gap which indicates an
almost uniform probability for ionization across the entire gap. No significant effect on the primary
ionization properties has been found for varying the SF$_{6}$ fraction from 0.0\% to 0.5\%.
\\
In simulating the detector response, it is important to carry out a precise computation of the field 
configuration within the detector because it plays the principal role in the detector dynamics.
The detail electrostatic field map within a 30 cm $\times$ 30 cm bakelite RPC has been calculated
using two numerical methods, namely, finite element method (COMSOL Multiphysics$^{\tiny{\textregistered}}$~\cite{comsol}) 
and boundary element method (neBEM) and the results are available
in~\cite{paper1}. For the present work, the field maps have been calculated using neBEM v1.8.20. 
A relative dielectric constant of 5.4 has been used for calculating the field in the bakelite RPC. 
The same calculation has been repeated for a glass RPC using a different relative dielectric constant
of 13.5, for the glass electrodes in geometry. The time dependence of the electric
field owing to the finite bulk resistivity of the RPC plate has been ignored as it is not expected to
affect the signal~\cite{riegler_finiteResistivity}. An appropriate approach for this study would be 
to find out the signal in the presence of a dynamic electric field  as the space charges created in
the avalanche process tend to modify it~\cite{lippmann_spaceCharge}. However, the calculations
in the present work have been performed assuming a static electric field configuration where the
RPC is described as a multi-dielectric planar capacitor.
\\
The electron transport properties in different gas mixtures have been calculated using the standalone 
Magboltz v8.9.3 program for different field values. The variation of the electron transport properties,
such as, effective Townsend coefficient ($\alpha_{eff} = \alpha - \eta$) which is the value after 
subtracting the attachment coefficient ($\eta$) from the first Townsend coefficient ($\alpha$), 
drift velocity (V$_{z}$) and longitudinal (D$_{l}$) and transverse (D$_{t}$) diffusion coefficients
have been calculated for C$_{2}$H$_{2}$F$_{4}$ based gas mixtures containing 5\% i-C$_{4}$H$_{10}$ 
and a small fraction of SF$_{6}$ that varies between 0.0 - 0.5\%. 
\begin{figure}[!htb]
 \centering
  \subfigure[]{
    \includegraphics[width=0.42\textwidth, trim={0cm 0.1cm 2.2cm 2.0cm},clip]{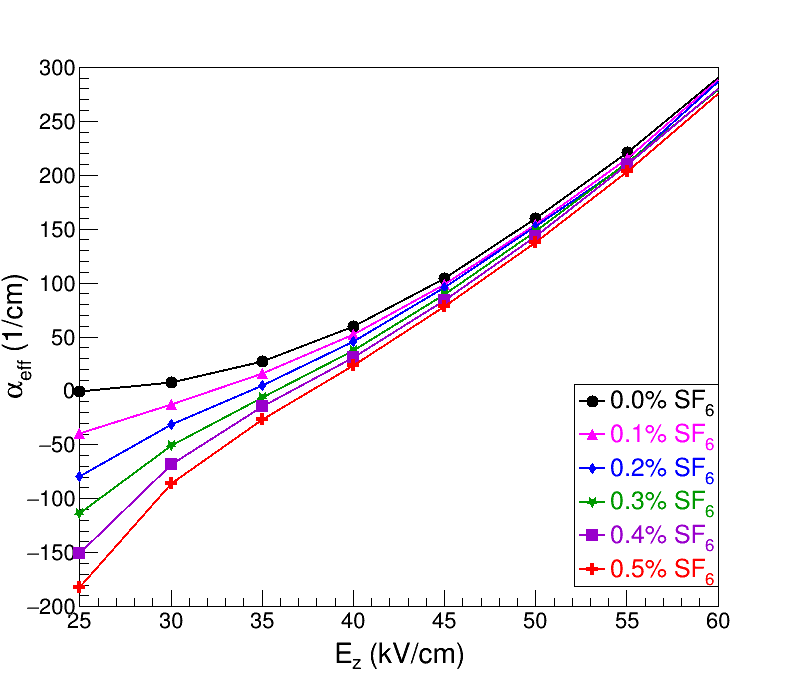}
  \label{fig:alphaEff_vs_Ez}
  }
  \subfigure[]{
     \includegraphics[width=0.42\textwidth, trim={0cm 0.1cm 2.2cm 2.2cm},clip]{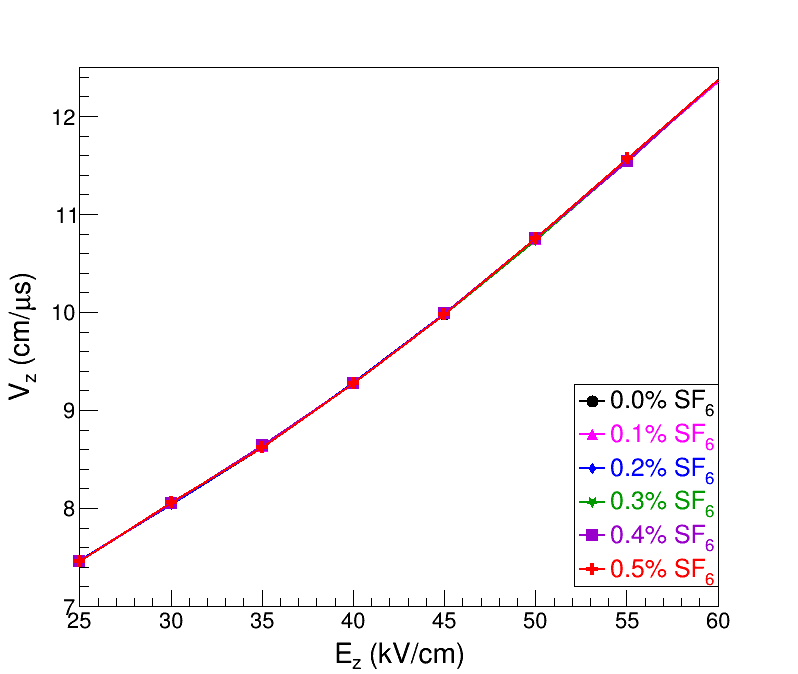}
\label{fig:Vz_vs_Ez}
  }
\caption{Variation of \subref{fig:alphaEff_vs_Ez} effective Townsend coefficient ($\alpha_{eff}$) 
and \subref{fig:Vz_vs_Ez} drift velocity of electrons (V$_{z}$) with the applied field ($E_{z}$) 
for different fractions of SF$_{6}$ in the gas mixture C$_{2}$H$_{2}$F$_{4}$ + 5\% i-C$_{4}$H$_{10}$ 
+ SF$_{6}$. Error bars within marker size.}
\label{fig:alphaEff_Vz_vs_Ez}
\end{figure}
The results are shown in figure~\ref{fig:alphaEff_Vz_vs_Ez} and~\ref{fig:Dl_Dt_vs_Ez}.
\begin{figure}[!htb]
 \centering
  \subfigure[]{
    \includegraphics[width=0.42\textwidth, trim={0.1cm 0.1cm 1.8cm 1.5cm},clip]{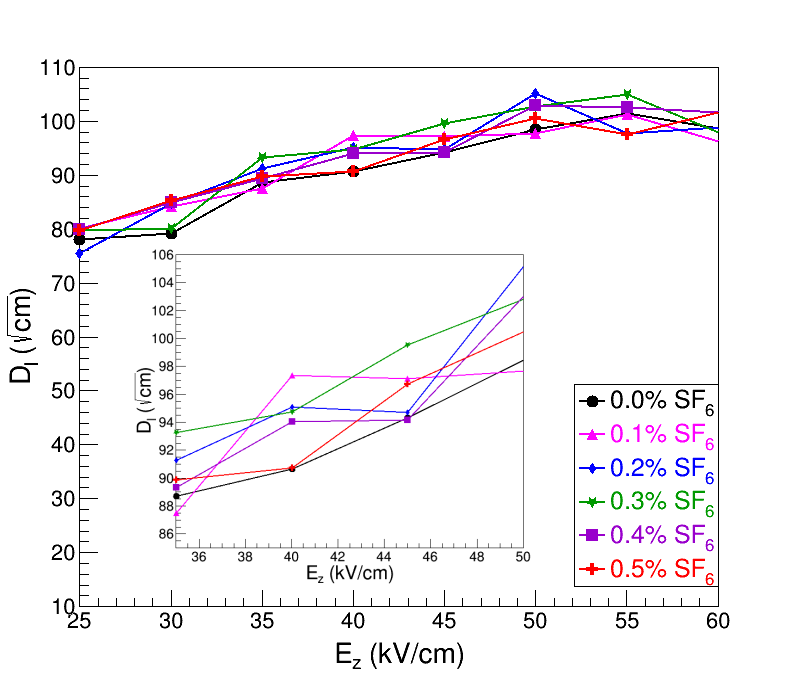}
  \label{fig:Dl_vs_Ez}
  }
  \subfigure[]{
     \includegraphics[width=0.42\textwidth, trim={0.1cm 0.1cm 1.8cm 1.5cm},clip]{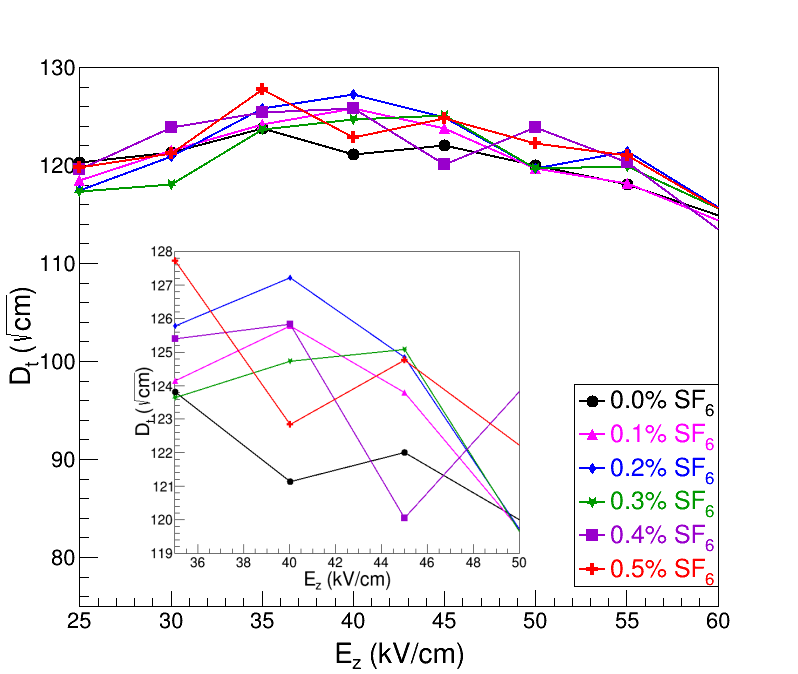}
\label{fig:Dt_vs_Ez}
  }
\caption{Variation of \subref{fig:Dl_vs_Ez} longitudinal ($D_{l}$) and \subref{fig:Dt_vs_Ez} transverse 
(D$_{t}$) diffusion coefficients of electrons with the applied field ($E_{z}$) for different fractions 
of SF$_{6}$ in the gas mixture C$_{2}$H$_{2}$F$_{4}$ + 5\% i-C$_{4}$H$_{10}$ + SF$_{6}$
(insets of both figures contain the variations in a shorter scale). Error bars within marker size.}
\label{fig:Dl_Dt_vs_Ez}
\end{figure}
It can be seen from figure~\ref{fig:alphaEff_vs_Ez} that with the increase in applied field, the effective 
Townsend coefficient increases signifying more ionizations and less attachment of electrons for any 
given mixture. It reduces at a given field value when the amount of SF$_{6}$ is increased in the
mixture for the obvious reason that SF$_{6}$ has a great affinity of electrons. In figure~\ref{fig:Vz_vs_Ez}, 
the drift velocity of electrons increases with the field which makes the detector response faster.
However, there is no effect of changing SF$_{6}$ fraction observed on this parameter.
Figure~\ref{fig:Dl_vs_Ez} and~\ref{fig:Dt_vs_Ez} displays the variation of longitudinal and transverse
diffusion coefficients with the applied electric field. In case of longitudinal diffusion, a rising trend 
in the coefficient with the increase in the field has been observed. On the other hand, the transverse 
diffusion coefficient shows a rise upto a field value of about 40 kV/cm and a subsequent fall. The 
variation of SF$_{6}$ fraction in the gas mixture does not show any distinguishable effect
on any of the diffusion coefficients.
\\
The signal induced on a RPC read-out strip has been calculated by passing muons of energy randomly 
varying between 0.5 - 10 GeV through the detector in randomly varying directions restricting the
incidence angle ($\theta$) within 0$^{\circ}$ - 10$^{\circ}$. The movement of all the electrons
and ions created in the gas mixture through the avalanche process has been tracked by the 
Garfield and the current induced on the read-out strip due to their movement has been calculated
at an interval of 100 ps. 
A typical signal shape as found from numerical calculation is shown in figure~\ref{fig:simuSignal_timeThreshold}.
\begin{figure}[!htb]
 \centering
  \subfigure[]{
  \includegraphics[width=0.42\textwidth, trim={0cm 0.2cm 2.0cm 1.9cm},clip]{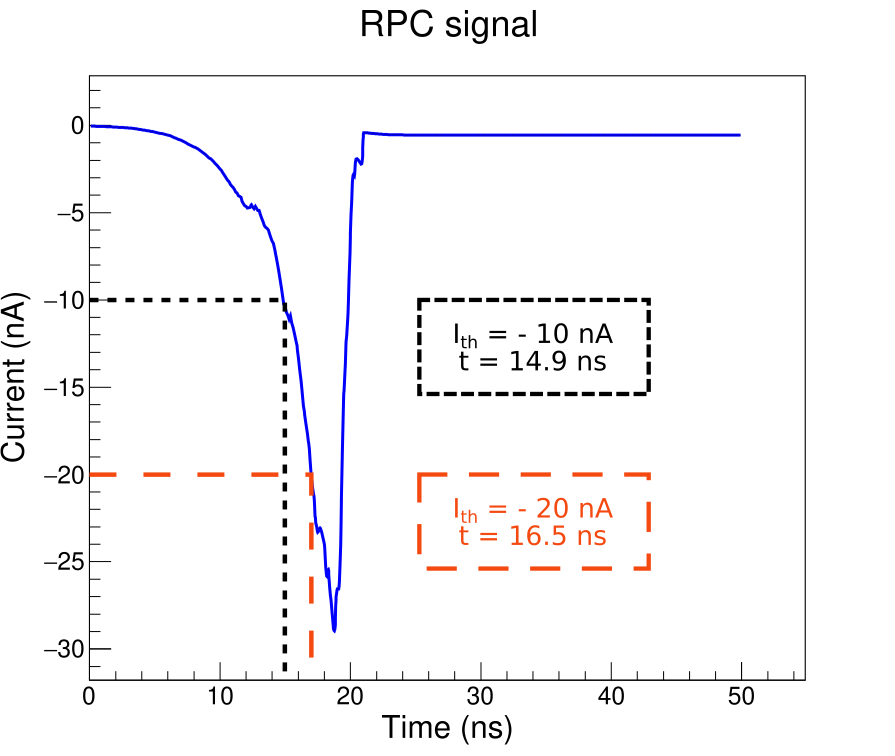}
  \label{fig:simuSignal_timeThreshold}
  }
  \subfigure[]{
   \includegraphics[width=0.42\textwidth, trim={0.cm 0.2cm 2.0cm 1.9cm},clip]{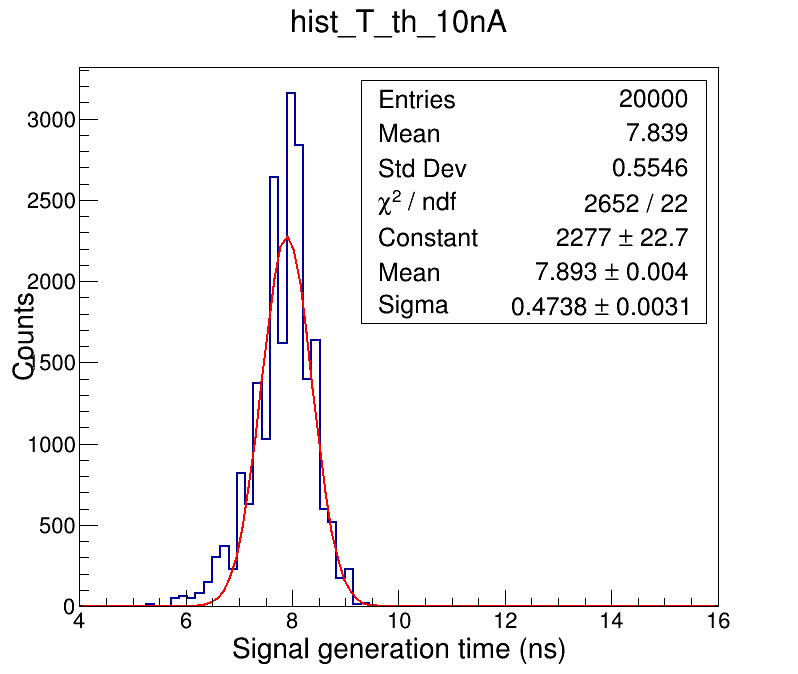}
  \label{fig:timingDist_simu}
  }
\caption{\subref{fig:simuSignal_timeThreshold} A typical simulated RPC signal and the scheme 
of selecting signal generation time corresponding to crossing different set thresholds, 
\subref{fig:timingDist_simu} distribution of signal generation times for 2000 events and its fit with 
Gaussian function for an RPC operated with C$_{2}$H$_{2}$F$_{4}$, 5\% i-C$_{4}$H$_{10}$ and 
0.2\% SF$_{6}$ at 42 kV/cm (used threshold = 10 nA).}
\label{fig:simu_signal_timingMethod}
\end{figure}
The experimentally obtained signal shapes generally appear with a long tail which is absent in the 
simulated current signal. This can be attributed to the assumption of simplified ion movement 
and non-inclusion of the effect of the electronics and the detector components outside the gas chamber.
The positive ions, having a larger mass and contributing to the slower falling edge, have been treated 
as moving with a constant velocity and not contributing to the growth of avalanche. 
The computed signal shapes have been analysed using Root data analysis framework~\cite{ROOT} 
to find the timing properties of the detector. A threshold in the current value has been used in the
simulation for eliminating the low amplitude noise signals as is usually done in experiments. The 
selection of times corresponding to crossing two fixed values of current thresholds is indicated 
in figure~\ref{fig:simuSignal_timeThreshold}. The time corresponding to crossing a specific threshold 
has been defined as the signal generation time. A distribution of the signal generation times has
been obtained for 2000 events where each event represents an averaging of current signals due to 
passage of 10 muons of randomly varying energy and angle of incidence within the mentioned range. A typical
distribution of signal generation times, to cross a threshold of 10 nA current, is shown in 
figure~\ref{fig:timingDist_simu}. The distribution has been fit with a Gaussian function to obtain the 
average signal generation time and the intrinsic time resolution of the detector from its mean and 
standard deviation, respectively. 
\\
The amplitude of the induced signals has been calculated for 2000 events (20000 muons with 
randomly varying energy and direction as mentioned earlier). A distribution of signal amplitudes
has been thus obtained for each set of applied field and gas mixture. 
\begin{figure}[!htb]
 \centering
  \subfigure[]{
   \includegraphics[width=0.42\textwidth, trim={0.2cm 0.2cm 2.5cm 2.2cm},clip]{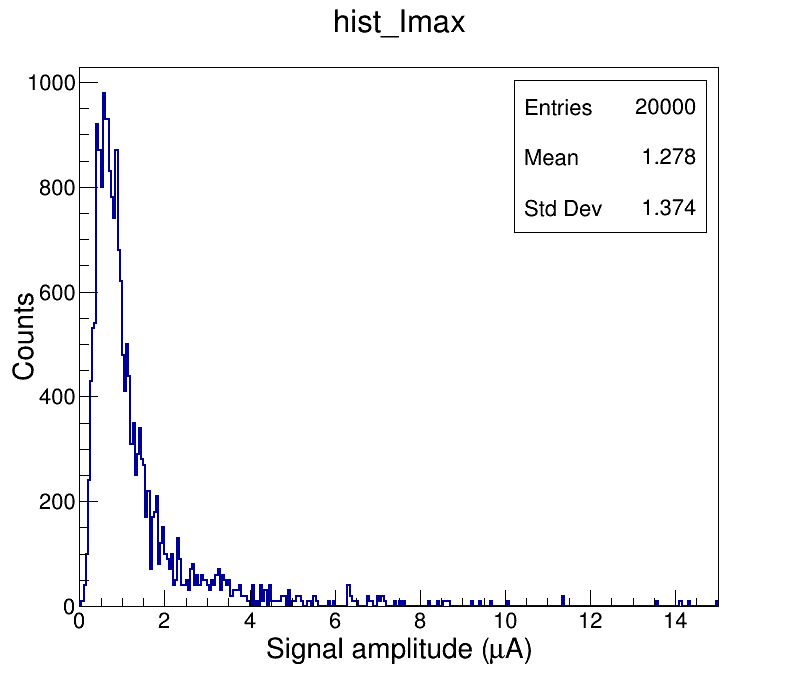}
  \label{fig:ImaxDist_simu}
  }
  \subfigure[]{
   \includegraphics[width=0.42\textwidth, trim={0.2cm 0.2cm 2.5cm 2.2cm},clip]{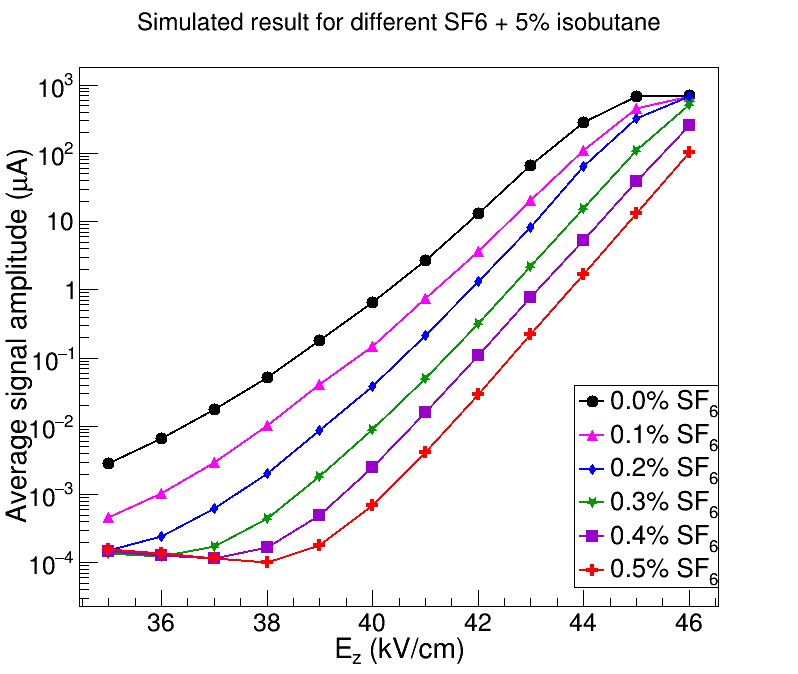}
  \label{fig:Imax_vs_HV_log}
  }
\caption{\subref{fig:ImaxDist_simu} Distribution of signal amplitudes from a RPC operated with 
C$_{2}$H$_{2}$F$_{4}$, 5\% i-C$_{4}$H$_{10}$ and 0.2\% SF$_{6}$ at 42 kV/cm, 
\subref{fig:Imax_vs_HV_log} variation of average signal amplitude with the applied field for 
different fractions of SF$_{6}$ in the gas mixture, containing 5\% i-C$_{4}$H$_{10}$ and rest 
C$_{2}$H$_{2}$F$_{4}$.}
\label{fig:simu_signal_timingMethod}
\end{figure}
One of such distributions 
when the RPC has been operated at a field of 42 kV/cm and with a gas mixture of 94.8\% 
C$_{2}$H$_{2}$F$_{4}$, 5\% i-C$_{4}$H$_{10}$ and 0.2\% SF$_{6}$ is shown in figure~\ref{fig:ImaxDist_simu}. 
The mean of the distribution, termed as average signal amplitude, has been used to denote the
characteristic response at different operating conditions.
To study the effect of the applied field and the SF$_{6}$ content in the gas mixture on the detector
efficiency, the average signal amplitudes have been calculated for different field values with a gas
mixture of C$_{2}$H$_{2}$F$_{4}$, 5\% i-C$_{4}$H$_{10}$ and different percentages of SF$_{6}$ 
varying from 0.0 to 0.5\%. The average signal amplitude were found to increase at higher electric 
field and for lower SF$_{6}$ in the gas mixture (figure~\ref{fig:Imax_vs_HV_log}). At higher fields
the rise in the Townsend coefficient or the gain in kinetic energy of the electrons caused release 
of larger number of electrons through further ionization of the gaseous molecules that contributed
to the rise in the induced current. On the other hand, the effect of addition of a trace amount of 
highly electronegative SF$_{6}$ in the gas mixture caused a reduction in the signal amplitude 
owing to attachment of electrons with the SF$_{6}$ molecules. Addition of SF$_{6}$ to the gas 
mixture is necessary to restrict streamer generation which may deteriorate the detector performance.
The detector efficiency has been calculated by finding out the fraction
of events which has been able to generate a signal amplitude more than a specified threshold.
\section{Comparison of numerical and experimental data}
\label{section:comp_simu_expt}
Experimental results for variation of the timing parameters with the fraction of SF$_{6}$
for a glass RPC of size 2 m $\times$ 2 m is available in~\cite{manas_glassRPCDevelopment}. 
Numerical calculations have been done for the same prototype operated with the same gas mixtures 
(C$_{2}$H$_{2}$F$_{4}$, 4.5\% i-C$_{4}$H$_{10}$ and SF$_{6}$) with smaller lateral 
dimension (16 cm $\times$ 16 cm) using the dielectric constant of Asahi glass~\cite{raveendra_glassProperties}
for the field calculation. The average signal generation time and intrinsic time resolution of the RPC
have been calculated at different applied fields.
\begin{figure}[!htb]
 \centering
  \subfigure[]{
   \includegraphics[width=0.42\textwidth, trim={0.1cm 0.1cm 2.5cm 2.2cm},clip]{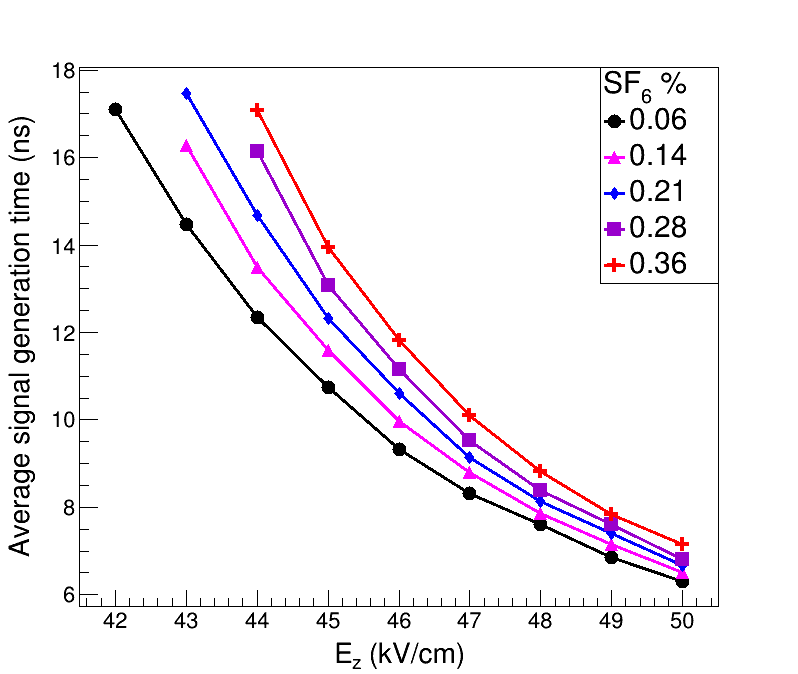}
  \label{fig:arrivalTime_vs_Ez-glassRPC}
  }
  \subfigure[]{
   \includegraphics[width=0.42\textwidth, trim={0.1cm 0.1cm 2.5cm 2.2cm},clip]{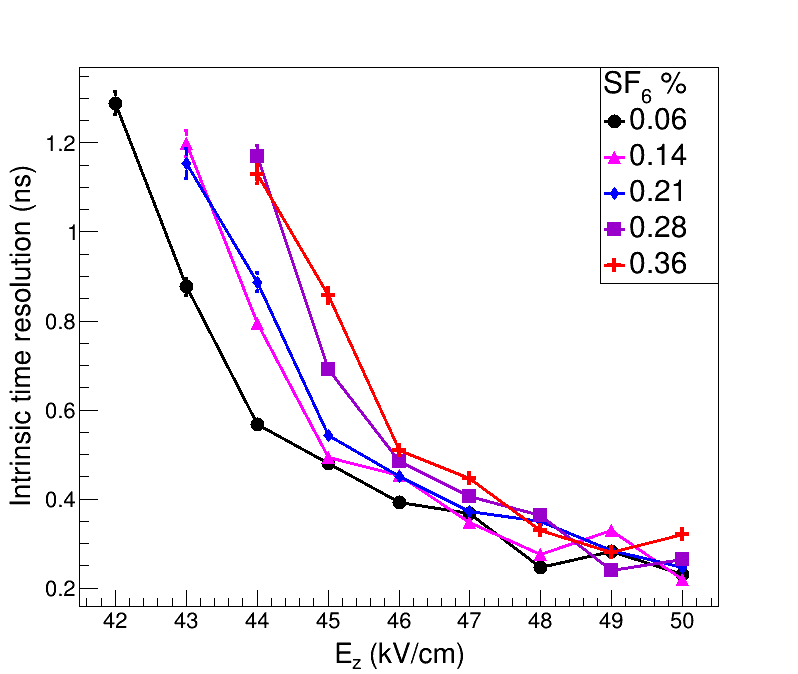}
  \label{fig:timeResolution_vs_Ez-glassRPC}
  }
\caption{Variation of \subref{fig:arrivalTime_vs_Ez-glassRPC} average signal generation time in 
crossing 2 $\mu$A threshold, and \subref{fig:timeResolution_vs_Ez-glassRPC} intrinsic time resolution
at the same condition with the applied field for a glass RPC operated with gas mixtures containing
different fractions of SF$_{6}$ in the mixture C$_{2}$H$_{2}$F$_{4}$ + i-C$_{4}$H$_{10}$ (4.5\%)
+ SF$_{6}$.}
\label{fig:simu_timing_vs_Ez-glassRPC}
\end{figure}
The variation of the two parameters, calculated for 2 $\mu$A threshold (corresponding voltage
threshold = 10 mV), with the applied field is shown in figure~\ref{fig:simu_timing_vs_Ez-glassRPC}
for different gas mixtures containing different fractions of SF$_{6}$ along with C$_{2}$H$_{2}$F$_{4}$
and 4.5\% i-C$_{4}$H$_{10}$.
\\
In the experiments~\cite{manas_glassRPCDevelopment}, both the timing parameters were found 
to deteriorate with the increase in SF$_{6}$ fraction in the gas mixture. The variation of the two 
timing parameters with the fraction of SF$_{6}$ has been found out from the numerical data at 
E$_{z}$ = 45 kV/cm, and compared with the experimental result in figure~\ref{fig:simu_timing_vs_SF6-glassRPC}.
For the comparison, the simulated values of arrival time have been 
incremented by 47.85 ns, as a contribution from the delay introduced by the electronic modules and
wires in the experiment. Similarly, a value of 1.33 ns has been added in quadrature to the simulated 
values of time resolution to take care of the jitters introduced by the related components. These 
two values are chosen arbitrarily to obtain the best match between the two curves.
\begin{figure}[!htb]
 \centering
  \subfigure[]{
   \includegraphics[width=0.42\textwidth, trim={0.1cm 0.1cm 2.5cm 2.0cm},clip]{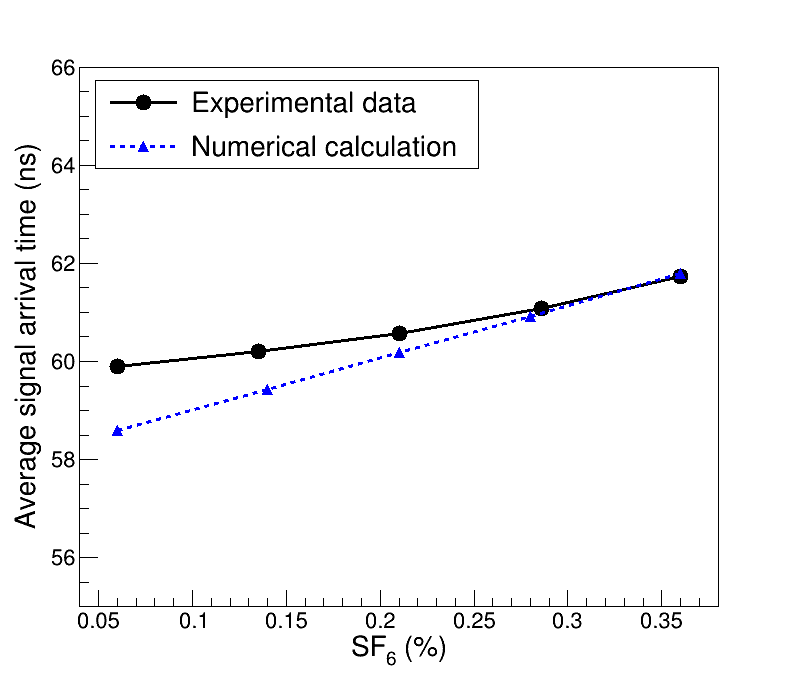}
  \label{fig:arrivalTime_vs_SF6-glassRPC}
  }
  \subfigure[]{
   \includegraphics[width=0.42\textwidth, trim={0.1cm 0.1cm 2.5cm 2.0cm},clip]{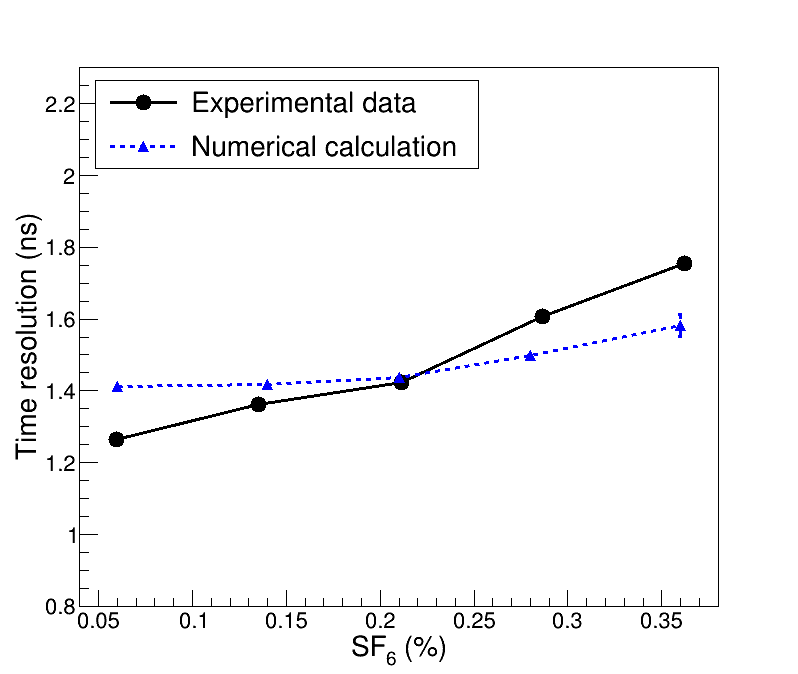}
  \label{fig:timeResolution_vs_SF6-glassRPC}
  }
\caption{Comparison of experimental and simulated results for the variation of 
\subref{fig:arrivalTime_vs_SF6-glassRPC} average signal arrival time and 
\subref{fig:timeResolution_vs_SF6-glassRPC} time resolution of a glass RPC with the fraction of
SF$_{6}$ present in the gas mixture (experimental data from~\cite{manas_glassRPCDevelopment}, 
simulation results at E$_{z}$ = 45 kV/cm).}
\label{fig:simu_timing_vs_SF6-glassRPC}
\end{figure}
The mismatch of arrival times for lower SF$_{6}$ fractions in the gas mixture (figure~\ref{fig:arrivalTime_vs_SF6-glassRPC})
can be attributed to the space charge generated within RPC in the experiment which slows it down 
by reducing the local electric field. No such effect is present in the simulation and thus is expected 
to produce a smaller value of signal arrival time compared to the experiment. Smaller the
fraction of SF$_{6}$ in the gas mixture larger is the charge production giving rise to more difference 
between the two curves. 
Although both simulation and experiment showed that the time resolution worsens with the increase 
in SF$_{6}$ fraction (figure~\ref{fig:timeResolution_vs_SF6-glassRPC}), the two curves were found to differ slightly in their nature.
\section{Numerical results}
\label{section:numericalResult}
\subsection{Effect of operational parameters}
\label{section:effect_operationalParameters}
Figure~\ref{fig:simu_timing_vs_Ez} shows the variation of average signal generation time and intrinsic 
time resolution of a bakelite RPC with the applied field when the RPC is operated with gas mixtures containing 
different fractions of SF$_{6}$ along with C$_{2}$H$_{2}$F$_{4}$ and 5\% i-C$_{4}$H$_{10}$.
\begin{figure}[!htb]
 \centering
  \subfigure[]{
   \includegraphics[width=0.45\textwidth, trim={0.2cm 0.2cm 2.5cm 2.2cm},clip]{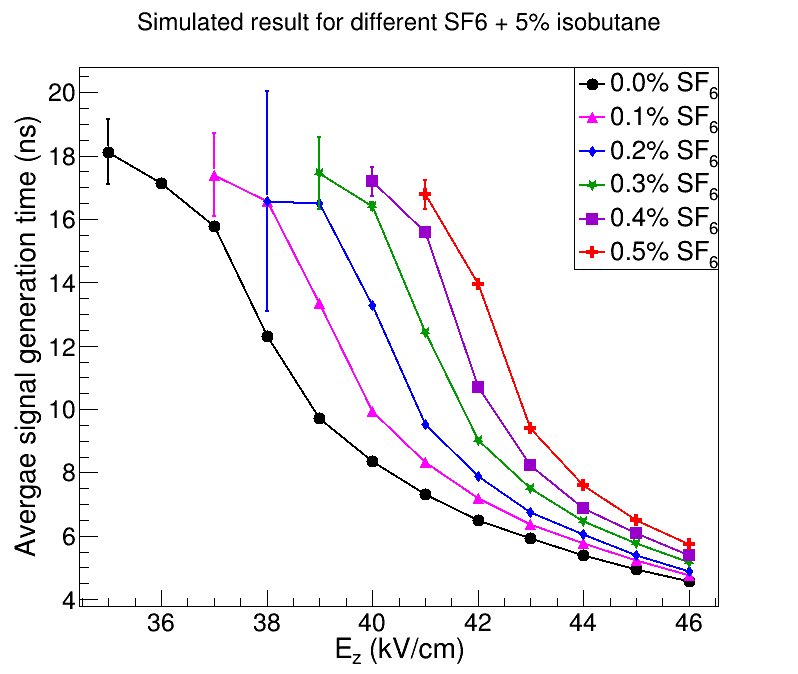}
  \label{fig:arrivalTime_vs_Ez}
  }
  \subfigure[]{
   \includegraphics[width=0.45\textwidth, trim={0.2cm 0.2cm 2.5cm 2.2cm},clip]{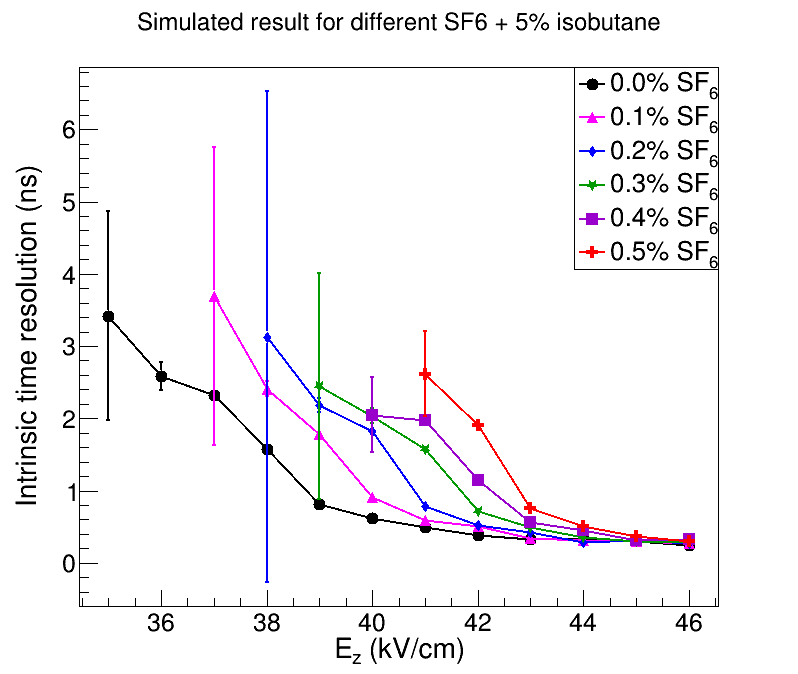}
  \label{fig:timeResolution_vs_Ez}
  }
\caption{Variation of \subref{fig:arrivalTime_vs_Ez} average signal generation time and 
\subref{fig:timeResolution_vs_Ez} intrinsic time resolution of a bakelite RPC with the applied field for 
different fraction of SF$_{6}$ content in the used gas mixture (C$_{2}$H$_{2}$F$_{4}$ + 5\% i-C$_{4}$H$_{10}$ 
+ SF$_{6}$) when the set threshold is 10 nA.}
\label{fig:simu_timing_vs_Ez}
\end{figure}
The curves have been obtained by setting a threshold of 10 nA. Values of both the timing 
parameters have been found to decrease at higher fields implying a better timing performance
of the detector. At higher values of field avalanche multiplication is larger which produces the 
detectable signal earlier and in turn reduces the average signal generation time; the increase of both
the effective Townsend coefficient and drift velocity of electrons at higher field region (refer to 
figure~\ref{fig:alphaEff_Vz_vs_Ez}) is responsible for the improvement of the time resolution 
according to the analytic formula described in~\cite{riegler_timingAnalytic}. The effect of 
SF$_{6}$ on the timing properties is also evident 
from figure~\ref{fig:simu_timing_vs_Ez}. The timing performance deteriorates for the gas mixtures
containing higher fraction of SF$_{6}$. The difference in the values for the two timing parameters
for different gas mixtures is prominent at lower field values which diminishes at higher values of 
the field. 
\begin{figure}[!htb]
 \centering
  \subfigure[]{
   \includegraphics[width=0.42\textwidth, trim={0.2cm 0.2cm 2.2cm 1.8cm},clip]{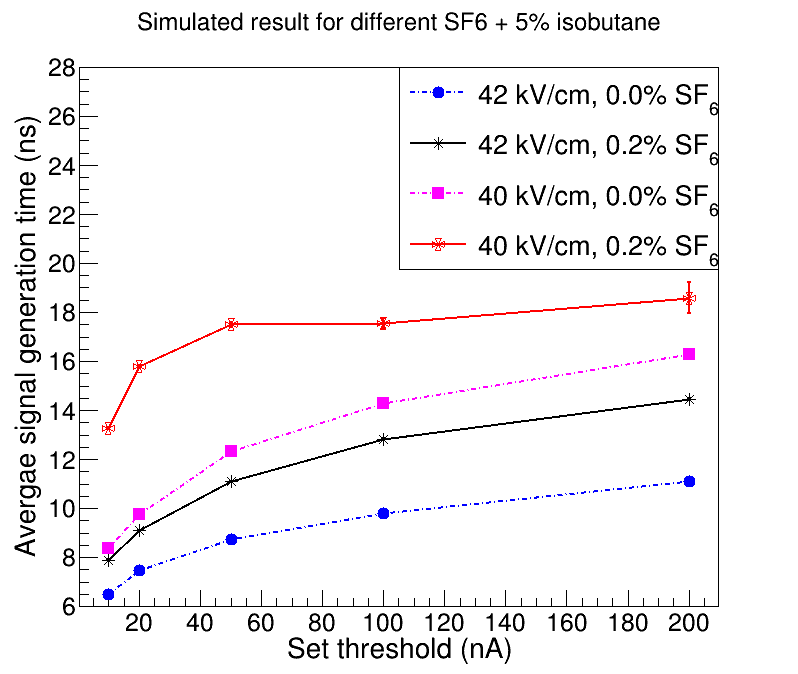}
  \label{fig:arrivalTime_vs_Ith}
  }
  \subfigure[]{
   \includegraphics[width=0.42\textwidth, trim={0.2cm 0.2cm 2.2cm 1.8cm},clip]{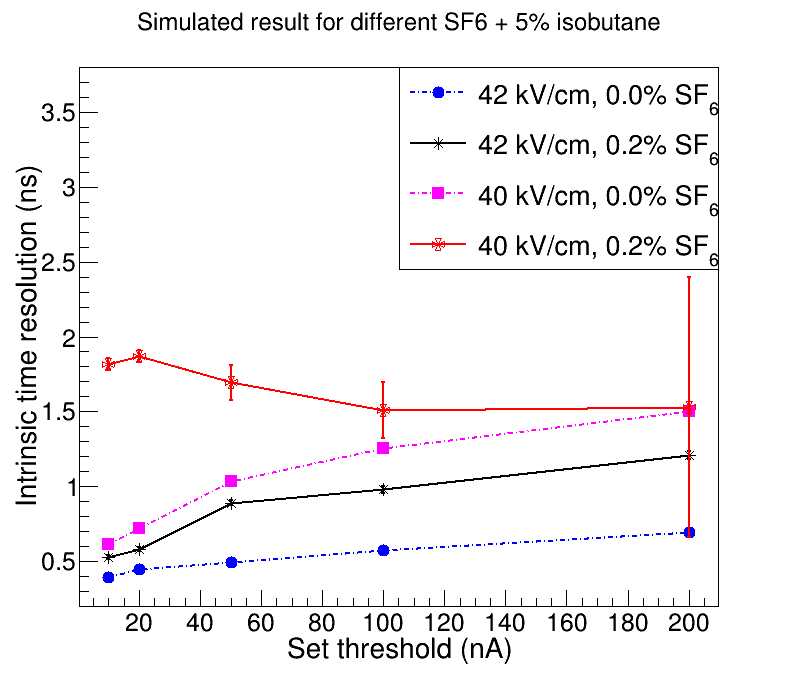}
  \label{fig:timeResolution_vs_Ith}
  }
\caption{Variation of \subref{fig:arrivalTime_vs_Ith} average signal generation time and 
\subref{fig:timeResolution_vs_Ith} intrinsic time resolution of a bakelite RPC with the used 
threshold for different gas mixtures operated in presence of different field values.}
\label{fig:simu_timing_vs_Ith}
\end{figure}
\\
The value of the timing parameters has been found to depend on the used threshold also. 
Variation of the two timing parameters with the used threshold has been shown in figure~\ref{fig:simu_timing_vs_Ith} 
for two different fields and gas mixtures containing different fractions
of SF$_{6}$. Again in this case, the dependence of the timing parameters on the used threshold 
has been found to be more prominent for lower field values and with higher fraction of SF$_{6}$
in the gas mixture.
\\
The variation of the detector efficiency with the applied field for different fractions of SF$_{6}$ in 
the gas mixture is shown in figure~\ref{fig:simu_efficiency_vs_Ez_manySF6_Ith_10nA} for crossing
a threshold of 10 nA .
\begin{figure}[!htb]
 \centering
  \subfigure[]{
   \includegraphics[width=0.47\textwidth, trim={0.2cm 0.2cm 2.5cm 2.2cm},clip]{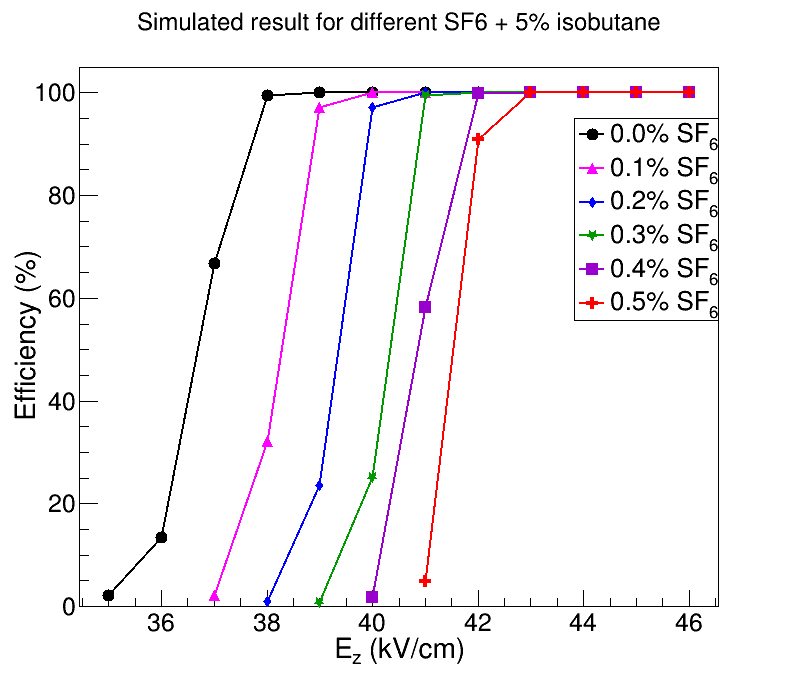}
  \label{fig:simu_efficiency_vs_Ez_manySF6_Ith_10nA}
  }
  \subfigure[]{
   \includegraphics[width=0.47\textwidth, trim={0.2cm 0.2cm 2.5cm 2.2cm},clip]{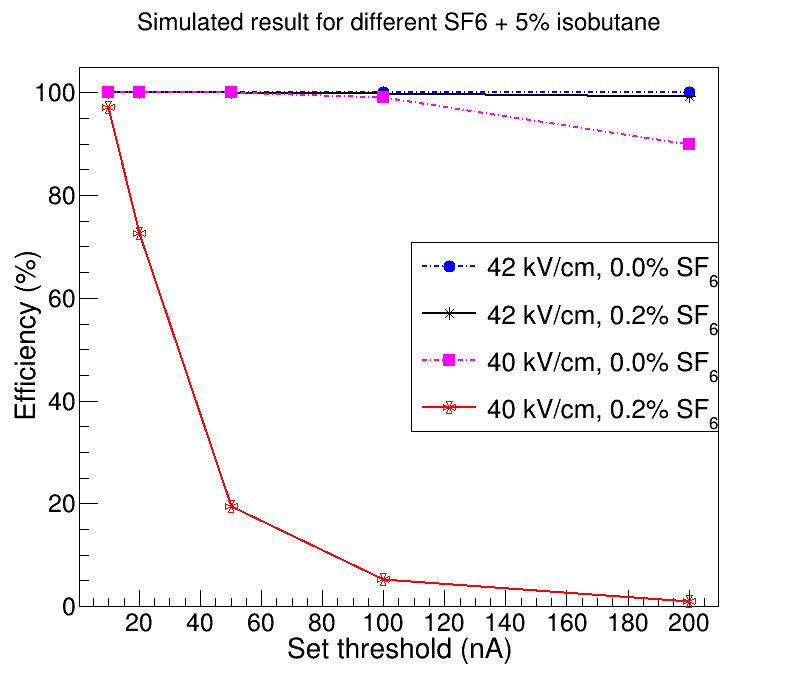}
  \label{fig:simu_efficiency_vs_Ith_many_Ez_SF6}
  }
\caption{Variation of RPC efficiency with \subref{fig:simu_efficiency_vs_Ez_manySF6_Ith_10nA} the applied 
field for different fractions of SF$_{6}$ in the gas mixture, containing 5\% i-C$_{4}$H$_{10}$ 
and rest C$_{2}$H$_{2}$F$_{4}$ (set threshold = 10 nA), \subref{fig:simu_efficiency_vs_Ith_many_Ez_SF6} the set 
threshold for two different fields for the gas mixtures containing C$_{2}$H$_{2}$F$_{4}$, 
5\% i-C$_{4}$H$_{10}$ and variable fractions of SF$_{6}$.}
\label{fig:simu_signalAmplitude_efficiency}
\end{figure}
The efficiency increased with the applied field and reached a plateau 
(100\% efficiency) after a while. For higher fraction of SF$_{6}$ in the gas mixture, a comparatively
higher value of field is required to reach the plateau region. The effect of used set threshold on
the detector efficiency can be found out from figure~\ref{fig:simu_efficiency_vs_Ith_many_Ez_SF6} 
where variation of detector efficiency with the used threshold has been shown for two different field 
values and two different fractions of SF$_{6}$ in the gas mixture. 
For lower values of field and moderate amount of SF$_{6}$, the detector efficiency
has been found to get reduced at higher threshold values as some of the events could not produce
signals crossing the set threshold. For higher field values and without inclusion of SF$_{6}$ the 
signal production is very efficient and the highest efficiency is achievable even when using a very 
high threshold ($\sim$ 200 nA).
In experiments, a minimum field (or applied voltage) is required to have a signal amplitude crossing the 
discriminator threshold and producing a non-zero efficiency, which increases with the increase in
the SF$_{6}$ fraction in the gas mixture.
\subsection{Effect of geometrical parameters}
\label{section:RPCresponse_EBSpacer}
Due to the distorted field map near the edge, and button spacers~\cite{paper1} of the RPC, 
the transport properties of the electrons are different at those regions which in turn influences 
the signal generation and timing properties of the detector at those regions. Muons of varying
energy and direction have been passed through the RPC at different distances from the edge 
spacer and button spacer of the typical shape to find their effect on the timing properties.
The variation of average signal generation time and the intrinsic time resolution at different distances 
from the edge spacer is shown in figure~\ref{fig:simu_timing_nearEdge} for the application of two 
\begin{figure}[!htb]
 \centering
  \subfigure[]{
   \includegraphics[width=0.42\textwidth, trim={0.2cm 0cm 2cm 2.0cm},clip]{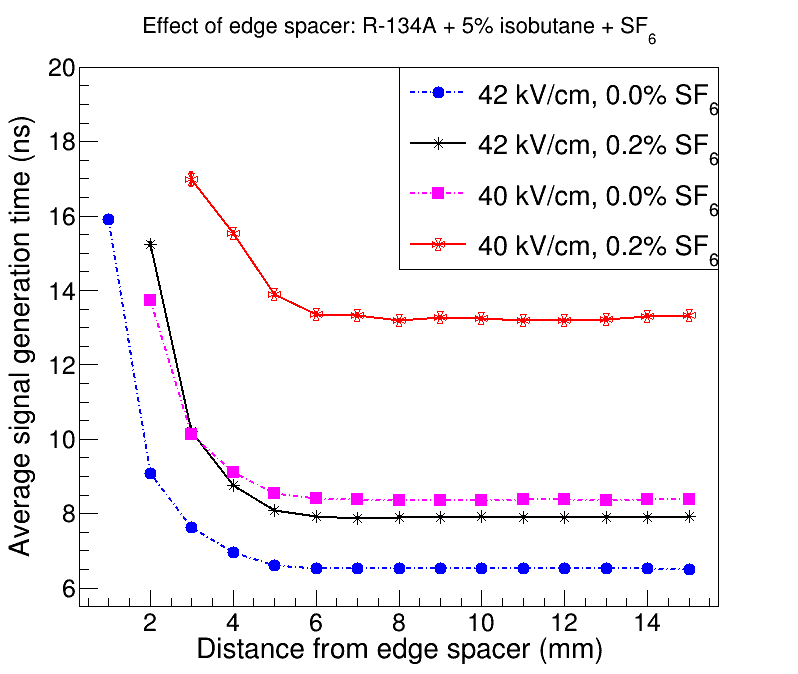}
  \label{fig:arrivalTime_nearEdge}
  }
  \subfigure[]{
    \includegraphics[width=0.42\textwidth, trim={0.2cm 0cm 2cm 2.0cm},clip]{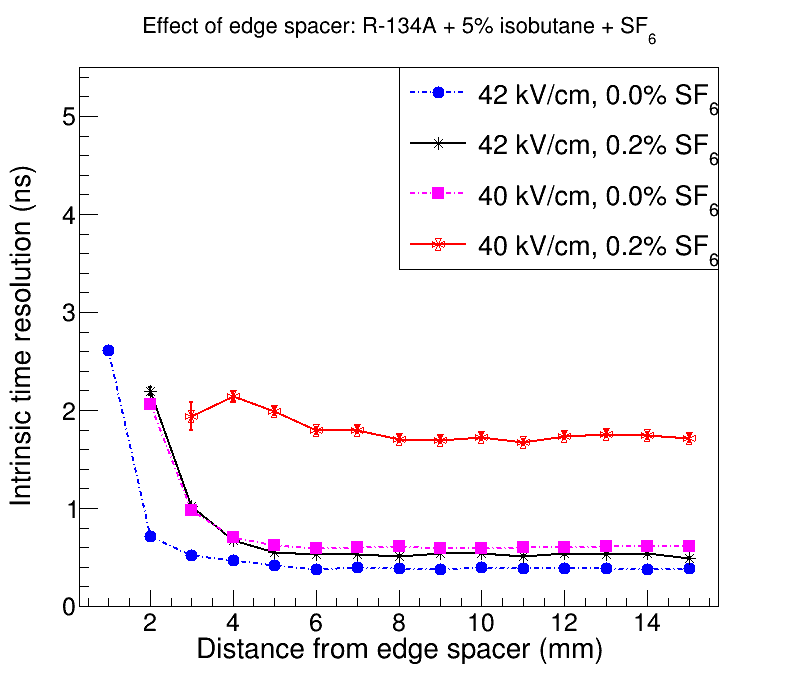}
  \label{fig:timeResolution_nearEdge}
  }
\caption{Variation of \subref{fig:arrivalTime_nearEdge} average signal generation time and 
\subref{fig:timeResolution_nearEdge} intrinsic time resolution of a bakelite RPC near the edge spacer
for different values of applied fields and different fractions of SF$_{6}$ in the gas mixture (used threshold = 10 nA).}
\label{fig:simu_timing_nearEdge}
\end{figure}
\begin{figure}[!htb]
 \centering
  \subfigure[]{
   \includegraphics[width=0.42\textwidth, trim={0.2cm 0cm 2cm 2.0cm},clip]{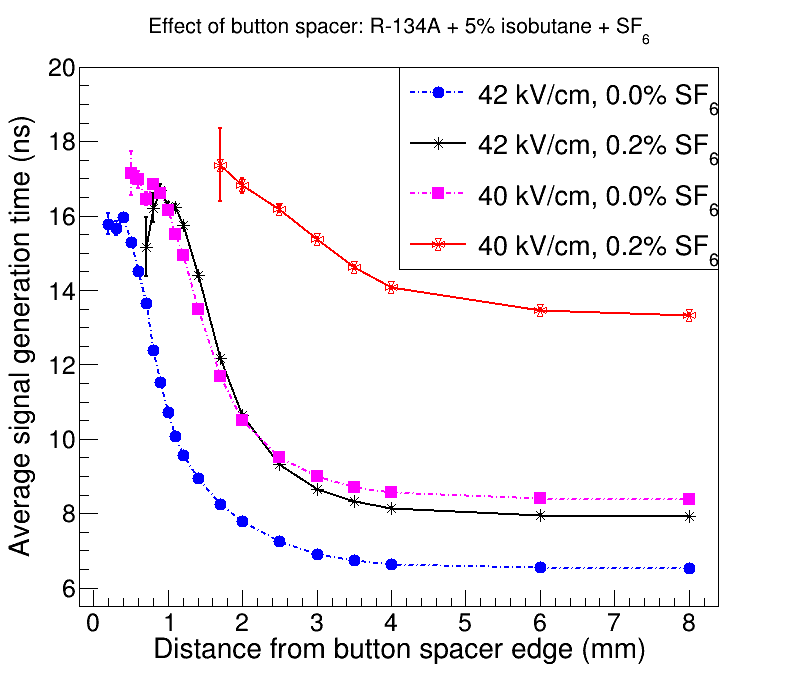}
  \label{fig:arrivalTime_nearBSpacer}
  }
  \subfigure[]{
   \includegraphics[width=0.42\textwidth, trim={0.2cm 0cm 2cm 2.0cm},clip]{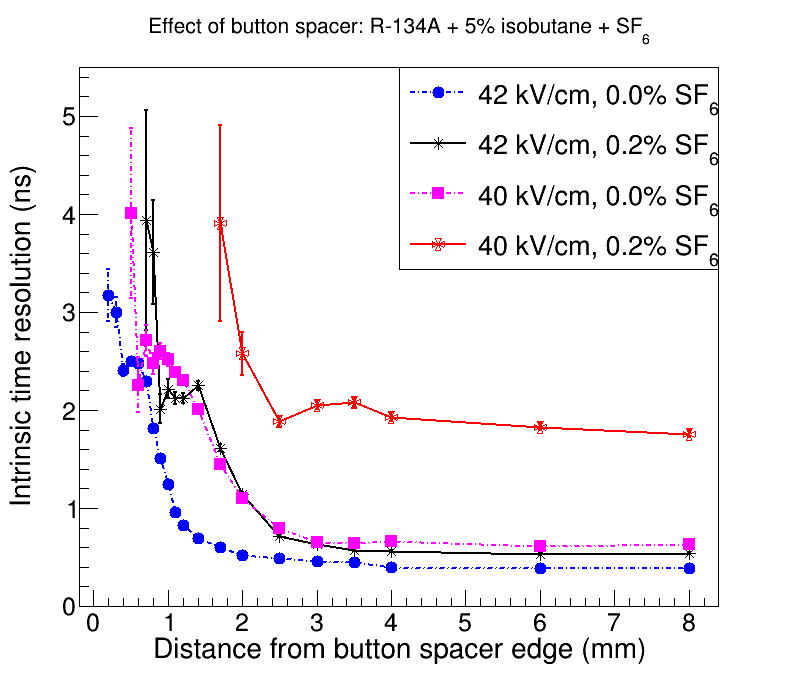}
  \label{fig:timeResolution_nearBSpacer}
  }
\caption{Variation of \subref{fig:arrivalTime_nearBSpacer} average signal generation time and 
\subref{fig:timeResolution_nearBSpacer} intrinsic time resolution of a bakelite RPC near a button 
spacer for different values of applied fields and different fractions of SF$_{6}$ in the gas mixture (used threshold = 10 nA).}
\label{fig:simu_timing_nearBSpacer}
\end{figure}
different fields and for gas mixtures containing two different fractions of SF$_{6}$ with 5\% i-C$_{4}$H$_{10}$
and rest C$_{2}$H$_{2}$F$_{4}$.
presence of large errors on the data points close to the edge and button spacers are due to less 
statistics as the signals from those regions are of less amplitude and very few of them are able to cross
the set threshold.
It is evident from figure~\ref{fig:simu_timing_nearEdge} that the timing properties deteriorate
very near to the edge spacer and the effect extends upto 5 mm from it.
The variation of the same at different distances away from a button spacer is shown in 
figure~\ref{fig:simu_timing_nearBSpacer}. 
The button has a groove like structure, where the electric field is highly non-uniform and its value 
changes rapidly at the points within this non-uniform region~\cite{paper1}. This gives rise to a 
non-monotonic variation of the timing parameters in its close vicinity.
From the figures~\ref{fig:simu_timing_vs_Ith},~\ref{fig:simu_timing_nearEdge} and~\ref{fig:simu_timing_nearBSpacer} it can be concluded that 
lower field and higher amount of SF$_{6}$ in the gas mixture worsen its timing performance.
\\
The detector efficiency has been calculated at different distances away from the edge and button 
spacer of RPC by passing muons through those locations. The variation of efficiency with the 
distance from the edge spacer is shown in figure~\ref{fig:efficiency_nearESpacer} for two different 
field values and different fractions of SF$_{6}$ in the gas mixture.
\begin{figure}[!htb]
 \centering
  \subfigure[]{
   \includegraphics[width=0.42\textwidth, trim={0.2cm 0cm 2cm 2.0cm},clip]{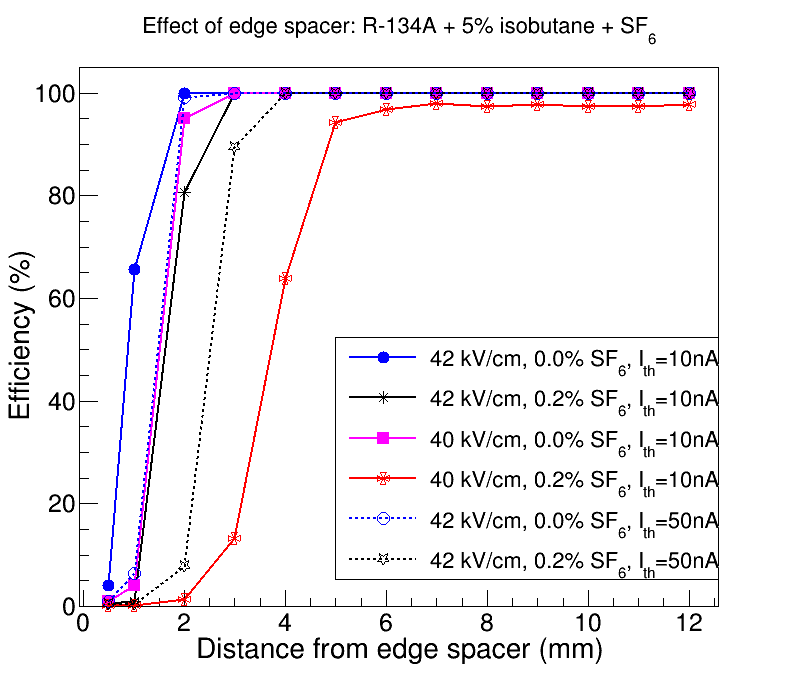}
  \label{fig:efficiency_nearESpacer}
  }
  \subfigure[]{
   \includegraphics[width=0.42\textwidth, trim={0.2cm 0cm 2cm 2.0cm},clip]{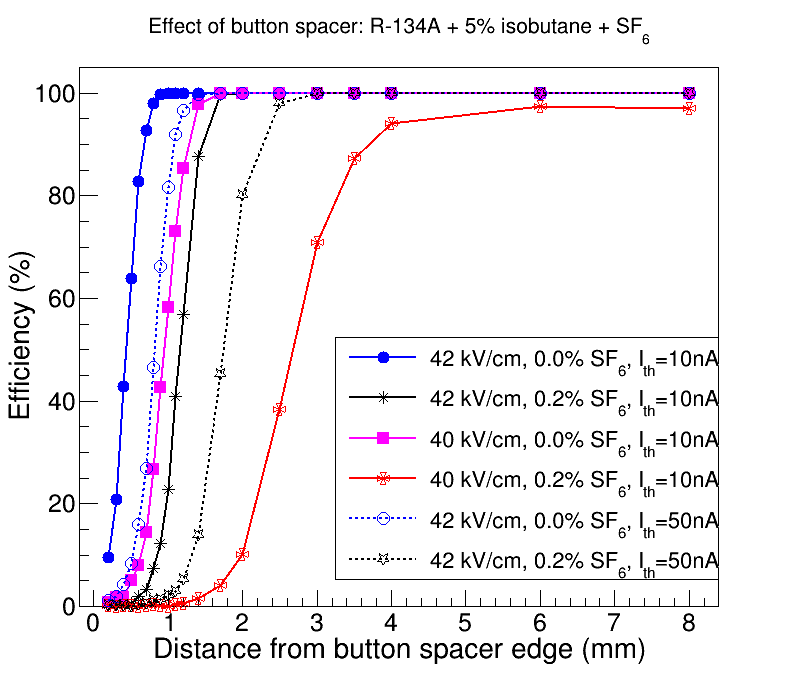}
  \label{fig:efficiency_nearBSpacer}
  }
\caption{Variation of detector efficiency with the \subref{fig:efficiency_nearESpacer} distance from 
edge spacer, \subref{fig:efficiency_nearBSpacer} distance from the edge of button spacer
for different values of applied fields (40 kV/cm, 42 kV/cm) and different fractions of SF$_{6}$ (0.0\%,
0.2\%) in the gas mixture and for different used thresholds (10 nA, 50 nA).}
\label{fig:simu_efficiency_nearEBSpacer}
\end{figure}
The results obtained using a threshold of 10 nA are shown in solid lines whereas that using a 
threshold of 50 nA are shown in dotted lines in the same figure. For a fixed threshold, presence 
of higher amount of SF$_{6}$ in the gas mixture reduces the signal amplitude. Also operating the 
detector at lower field value has the same effect. These in turn produces a lower detection efficiency.
Raising the value of used threshold decreases the number of events able to cross the threshold and 
in turn reduces the efficiency which is also visible in figure~\ref{fig:efficiency_nearESpacer} for 
two different combinations of field and gas mixture. The electric field suffers near the spacers which 
is the reason for lower signal production in their vicinity. The variation of the detector efficiency with 
the distance from button spacer edge is shown in figure~\ref{fig:efficiency_nearBSpacer} where 
a similar trend can be seen. The extent of the inefficient regions depend on the choice of applied 
field, used gas mixture and the set threshold. For the operation of the detector at 42 kV/cm with 0.2\%
SF$_{6}$, the inefficient region extends upto 3 mm from the edge spacer and about 1.75 mm from 
the button spacer if the used threshold is 10 nA. The inefficient region increases at higher threshold.
A surface map of the detector efficiency in X-Y plane of an RPC with all its components is shown in
figure~\ref{fig:simu_RPC_efficiency_map} where $X = 0$ and $Y = 0$ lines are the boundaries between
the gas chamber and the edge spacers. The map is generated for the RPC operated at 42 kV/cm
with the gas mixture containing 0.2\% SF$_{6}$ and for the threshold of 10 nA.
\begin{figure}[!htb]
	\centering
	\includegraphics[width=\textwidth, trim={1.0cm 0.2cm 0.2cm 0.2cm},clip]{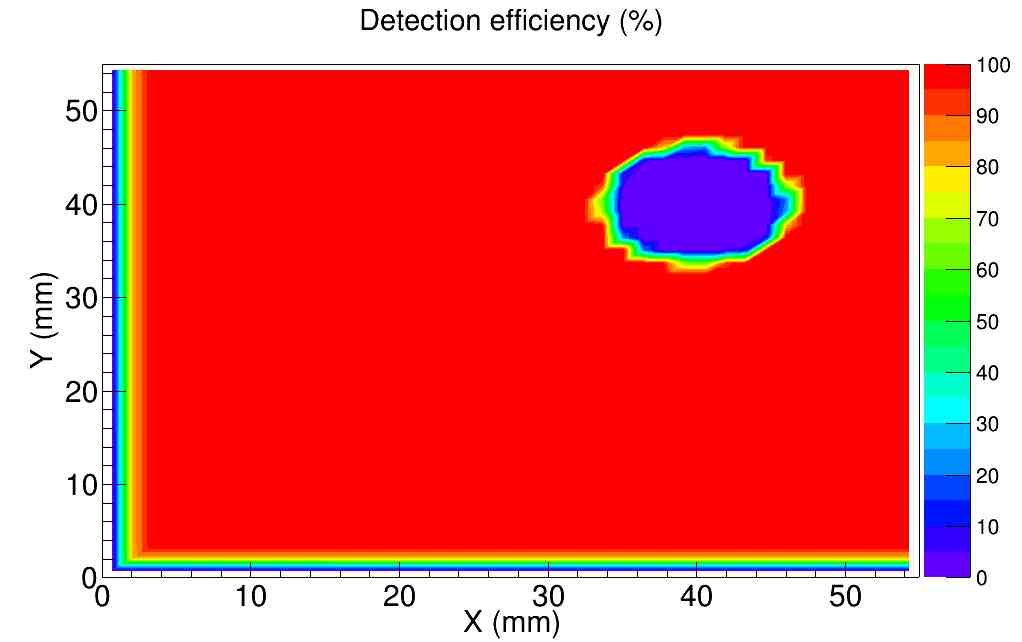}
	\caption{Surface map of RPC detection efficiency when it is operated with 94.8\% C$_{2}$H$_{2}$F$_{4}$,
	5\% i-C$_{4}$H$_{10}$ and 0.2\% SF$_{6}$ at 42 kV/cm (used threshold = 10 nA).}
	\label{fig:simu_RPC_efficiency_map}
\end{figure}
Effect of button spacer on the detection efficiency has been shown by placing a button spacer of
stem radius 5.1 mm and paddle radius 5.5 mm with its center at $X$ = $Y$ = 40 mm.
Similar kind of surface maps were found from experiments also~\cite{manas_RPCdeadSpace, apoorva_timeImprovement}.
\section{Computing resources}
\label{section:resources}
The calculations using Garfield have been performed in DELL PowerEdge R930 server having 
256 GB RAM, 64 cores, 2.2 GHz CPU running CentOS 7.5. For fixed operating conditions,
calculation of signals for the passage of 20000 muons, using field maps pre-calculated in neBEM
and gas files prepared by Magboltz stand alone separately, takes 1 day 5 hours to complete.
\section{Conclusion}
\label{section:Conclusion}
The amplitude of the RPC signal increases with the increase in the applied voltage which in turn 
improves the efficiency of the detector.
An RPC becomes faster with the increase in applied field (or voltage) and the time resolution also 
improves. However, raising the field to a very high value increases streamer probability~\cite{Cardarelli_avalanche2Streamer} 
which is known to affect the RPC timing performance adversely.
\\
The electron quenching role of SF$_{6}$ is evident from the simulated results. 
Although a certain amount of SF$_{6}$ is used to limit the streamer generation, the increase of 
SF$_{6}$ proportion in the gas mixture deteriorates the timing performance of the detector. 
So, to achieve the best timing performance in the avalanche mode operation with limited streamer
contribution, a RPC is required to be operated at an optimum voltage along with an optimum 
amount of SF$_{6}$ whose amount need to be chosen judiciously depending on the requirements
of the experiment.
\\
The simulation predicts that the response of RPC at critical regions like very near to edge and button
spacers gets altered due to the affected field map~\cite{paper1} at those regions. The timing response 
also gets affected adversely in those regions. The effect of these geometrical components vanish 
by about 7 mm away from edge spacers and about 5 mm away from the edge of the button spacers.
So, the effective volume of RPC producing uniform response will be little less than the geometrical
volume of its gas chamber. The detection efficiency also gets hampered near the spacers which in turn 
is expected to produce dead regions.
\\
The present robust simulation method can be used to find the timing performance of other gaseous 
detectors also, as well as their dependence on various operating and design parameters.
%
%
%
\section*{Acknowledgment}
The financial support and the helpful review received from INO collaboration is gratefully acknowledged.
A.~Jash is grateful to Mr.~Sridhar Tripathy and Mr.~Jaydeep Datta for their help and cooperation during this work.

\begin{thebibliography}{99}
%
\bibitem{RPC}
R. Santonico and R. Cardarelli,
\emph{Development of resistive plate counters},
\href{https://doi.org/10.1016/0029-554x(81)90363-3}{\emph{Nucl. Instrum. Meth. A}, {\bf 187} (1981) 377}.
%
\bibitem{ICAL}
ICAL collaboration,
\emph{Physics Potential of the ICAL detector at the India-based Neutrino Observatory (INO)},
\href{https://doi.org/10.1007/s12043-017-1373-4}{\emph{Pramana}, {\bf 88} (2017) 79} 
[\href{https://arxiv.org/abs/1505.07380}{arXiv:1505.07380}].
%
\bibitem{manas_glassRPCDevelopment}		
M. Bhuyan et al.,
\emph{Development of 2 m $\times$ 2 m size glass RPCs for INO},
\href{https://doi.org/10.1016/j.nima.2010.09.087}{\emph{Nucl. Instrum. Meth. A}, {\bf 661} (2012) S64}.
%
\bibitem{paper_Shockley}
W. Shockley,
\emph{Currents to conductors induced by a moving point charge},
\href{https://doi.org/10.1063/1.1710367}{\emph{J. of Appl. Phys.}, {\bf 9} (1938) 635}.
%
\bibitem{paper_Ramo}
S. Ramo,
\emph{Currents induced by electron motion},
\href{https://doi.org/10.1109/jrproc.1939.228757}{\emph{Proc. IRE}, {\bf 27} (1939) 584}.
%
\bibitem{Garfield}
R. Veenhoff,
\emph{GARFIELD, recent developments},
\href{https://doi.org/10.1016/S0168-9002(98)00851-1}{\emph{Nucl. Instrum. Meth. A}, {\bf 419} (1998) 726}
[\url{http://garfield.web.cern.ch/garfield}].
%
\bibitem{HEED}
I.B. Smirnov,
\emph{Modeling of ionization produced by fast charged particles in gases}, 
\href{https://doi.org/10.1016/j.nima.2005.08.064}{\emph{Nucl. Instrum. Meth. A}, {\bf 554} (2005) 474}
[\url{https://heed.web.cern.ch/heed/}].
%
\bibitem{neBEM}
N. Majumdar and S. Mukhopadhyay,
\emph{Simulation of 3D electrostatic configuration in gaseous detectors},
\href{https://doi.org/10.1088/1748-0221/2/09/P09006}{\emph{JINST}, {\bf 2} (2007) P09006}
[\url{http://nebem.web.cern.ch/nebem/}].
%
\bibitem{Magboltz}
S.F. Biagi,
\emph{Accurate solution of the Boltzmann transport equation},
\href{https://doi.org/10.1016/0168-9002(88)90050-2}{\emph{Nucl. Instrum. Meth. A}, {\bf 273} (1988) 533}
[\url{http://magboltz.web.cern.ch/magboltz/}].
%
\bibitem{comsol}
COMSOL Multiphysics$^{\tiny{\textregistered}}$ - multiphysics simulation tool, 
available at \href{www.comsol.com}{www.comsol.com}.
%
\bibitem{paper1}
A. Jash, N. Majumdar, S. Mukhopadhyay and S. Chattopadhyay,
\emph{Numerical studies on electrostatic field configuration of Resistive Plate Chambers for the INO-ICAL experiment},
\href{https://doi.org/10.1088/1748-0221/10/11/P11009}{\emph{JINST}, {\bf 10} (2015) P11009}.
%
\bibitem{riegler_finiteResistivity}
W. Riegler,
\emph{Induced signals in resistive plate chambers},
\href{https://doi.org/10.1016/S0168-9002(02)01169-5}{\emph{Nucl. Instrum. Meth. A}, {\bf 491} (2002) 258}.
%
\bibitem{lippmann_spaceCharge}
C. Lippmann and W. Riegler,
\emph{Space charge effects in Resistive Plate Chambers},
\href{https://doi.org/10.1016/j.nima.2003.08.174}{\emph{Nucl. Instrum. Meth. A}, {\bf 517} (2004) 54}.
%
\bibitem{ROOT}
R. Brun and F. Rademakers,
\emph{ROOT - An object oriented data analysis framework},
\href{https://doi.org/10.1016/S0168-9002(97)00048-X}{\emph{Nucl. Instrum. Meth. A}, {\bf 389} (1997) 81}
[\url{https://root.cern.ch/}].
%
\bibitem{raveendra_glassProperties}		
K. Raveendrababu, P. K. Behera, B. Satyanarayana, J. Sadiq,
\emph{Study of glass properties as electrode for RPC},
\href{https://doi.org/10.1088/1748-0221/11/07/C07007}{\emph{JINST}, {\bf 11} (2016) C07007}
[\href{https://arxiv.org/abs/1605.01044}{arXiv:1605.01044}].
%
\bibitem{riegler_timingAnalytic}
W. Riegler and C. Lippmann,
\emph{Detailed models for timing and efficiency in resistive plate chambers},
\href{https://doi.org/10.1016/S0168-9002(03)01269-5}{\emph{Nucl. Instrum Meth. A}, {\bf 508} (2003) 14}.
%
\bibitem{manas_RPCdeadSpace} 		
M. Bhuyan et al.,
\emph{Cosmic ray test of INO RPC stack},
Figure 9, \href{https://doi.org/10.1016/j.nima.2010.08.034}{\emph{Nucl. Instrum. Meth. A}, {\bf 661} (2012) S68}.
%
\bibitem{apoorva_timeImprovement}		
A. D. Bhatt, V. M. Datar, G. Majumdar, N. K. Mondal, Pathaleswar and B. Satyanarayana,
\emph{Improvement of time measurement with the INO-ICAL resistive plate chambers},
Figure 1(a,b), \href{https://doi.org/10.1088/1748-0221/11/11/C11001}{\emph{JINST}, {\bf 11} (2016) C11001}.
%
\bibitem{Cardarelli_avalanche2Streamer} 		
R. Cardarelli, V. Makeev, R. Santonico,
\emph{Avalanche and streamer mode operation of resistive plate chambers},
Figure 4, \href{https://doi.org/10.1016/S0168-9002(96)00811-X}{\emph{Nucl. Instrum. Meth. A}, {\bf 382} (1996) 470}.
%
%
\end{thebibliography}
\end{document}